\documentclass[%
 reprint,
superscriptaddress,
 amsmath,amssymb,
 aps,
]{revtex4-2}

\usepackage{graphicx}
\usepackage{dcolumn}
\usepackage{bm}
\usepackage[mathlines]{lineno}

\makeatletter
\def\bibsection{%
   \par
   \begingroup
    \baselineskip26\p@
    \bib@device{\hsize}{72\p@}%
   \endgroup
   \nobreak\@nobreaktrue
   \addvspace{19\p@}%
  }%
\makeatother

\usepackage{siunitx}
\usepackage{hyperref}
\hypersetup{
	colorlinks=true,
	linkcolor=blue,
	filecolor=blue,      
	urlcolor=blue,
	citecolor=blue,
}

\usepackage{tikz}
\usetikzlibrary{shapes.geometric, arrows,positioning,calc,patterns.meta}
\pgfdeclarelayer{background}
\pgfdeclarelayer{foreground}
\pgfsetlayers{background,main,foreground}

\definecolor{lightblue}{RGB}{135,206,250}
\definecolor{darkblue}{RGB}{31,119,180}

\pgfdeclarepattern{
  name=hatch,
  parameters={\hatchsize,\hatchangle,\hatchlinewidth},
  bottom left={\pgfpoint{-.1pt}{-.1pt}},
  top right={\pgfpoint{\hatchsize+.1pt}{\hatchsize+.1pt}},
  tile size={\pgfpoint{\hatchsize}{\hatchsize}},
  tile transformation={\pgftransformrotate{\hatchangle}},
  code={
    \pgfsetlinewidth{\hatchlinewidth}
    \pgfpathmoveto{\pgfpoint{-.1pt}{-.1pt}}
    \pgfpathlineto{\pgfpoint{\hatchsize+.1pt}{\hatchsize+.1pt}}
    \pgfpathmoveto{\pgfpoint{-.1pt}{\hatchsize+.1pt}}
    \pgfpathlineto{\pgfpoint{\hatchsize+.1pt}{-.1pt}}
    \pgfusepath{stroke}
  }
}

\tikzset{
  hatch size/.store in=\hatchsize,
  hatch angle/.store in=\hatchangle,
  hatch line width/.store in=\hatchlinewidth,
  hatch size=5pt,
  hatch angle=0pt,
  hatch line width=.5pt,
}

\tikzstyle{materia}=[draw, fill=white, text width=8.0em, text centered,
  minimum height=1.5em]
\tikzstyle{materiahighlight}=[draw=black!50, fill=black!5, text width=8.0em, text centered,
  minimum height=1.5em, line width=1pt, dashed]
\tikzstyle{materiastart}=[draw, fill=green!10, text width=8.0em, text centered,
  minimum height=1.5em]
\tikzstyle{materiaend}=[draw, fill=red!10, text width=8.0em, text centered,
  minimum height=1.5em]
\tikzstyle{materiahf}=[draw, fill=darkblue!50, text width=8.0em, text centered,
  minimum height=1.5em]
\tikzstyle{materiarm}=[draw, fill=lightblue!30, text width=8.0em, text centered,
  minimum height=1.5em]
\tikzstyle{etape} = [materia, minimum width=8em,
  minimum height=3em, rounded corners]
\tikzstyle{etapehighlight} = [materiahighlight, minimum width=8em,
  minimum height=3em, rounded corners]
\tikzstyle{etapestart} = [materiastart, minimum width=8em,
  minimum height=3em, rounded corners]
\tikzstyle{etapeend} = [materiaend, minimum width=8em,
  minimum height=3em, rounded corners]
\tikzstyle{etapehf} = [materiahf, minimum width=8em,
  minimum height=3em, rounded corners]
\tikzstyle{etapegrid} = [materia, minimum width=8em,
  minimum height=3em, rounded corners, pattern=hatch, pattern color=lightblue!70, hatch size=6pt, hatch angle=0, preaction={fill=lightblue!30}]
\tikzstyle{etaperm} = [materiarm, minimum width=8em,
  minimum height=3em, rounded corners]
\tikzstyle{texto} = [above, text width=6em, text centered]
\tikzstyle{linepart} = [draw, thick, color=black!50, -latex', dashed]
\tikzstyle{line} = [draw, thick, color=black!50, -latex']
\tikzstyle{ur}=[draw, text centered, minimum height=0.01em]
\tikzstyle{decision} = [diamond, minimum width=1em, minimum height=1em, text centered, draw=black, fill=black!30]
\tikzstyle{checktext} = [text width=5em, text centered]
\tikzstyle{arrow} = [thick,->,>=stealth]
\tikzstyle{headlessarrow} = [thick,>=stealth]

\newcommand{\background}[5]{%
  \begin{pgfonlayer}{background}
    \path (#1.west |- #2.north)+(-0.3,0.3) node (#5-a1) {};
    \path (#3.east |- #4.south)+(+0.3,-0.3) node (#5-a2) {};
    \path[fill=black!5,rounded corners, draw=black!50, dashed]
      (#5-a1) rectangle (#5-a2);
  \end{pgfonlayer}}

\newcommand{\bThetaPlasmaSpecies}{\ensuremath{B^{(p)}_{\theta}}}
\newcommand{\bThetaBeamSpecies}{\ensuremath{B^{(b)}_{\theta}}}

\begin{document}
\preprint{APS/123-QED}

\title{Gridless Quasistatic Model for Efficient Simulation of Plasma-based Accelerators}

\author{A. Ferran Pousa}
\email{angel.ferran.pousa@desy.de}
\affiliation{Deutsches Elektronen-Synchrotron DESY, Notkestr. 85, 22607 Hamburg, Germany}

\author{W. M. den Hertog}
\affiliation{Universidade de Santiago de Compostela, Santiago de Compostela, Spain}
\altaffiliation[Now at ]{UHasselt, Faculty of Sciences, Agoralaan, 3590 Diepenbeek, Belgium}

\author{S. Diederichs}
\affiliation{Deutsches Elektronen-Synchrotron DESY, Notkestr. 85, 22607 Hamburg, Germany}
\affiliation{CERN, Espl. des Particules 1, 1211 Geneva, Switzerland}

\author{A. Martinez de la Ossa}
\affiliation{Deutsches Elektronen-Synchrotron DESY, Notkestr. 85, 22607 Hamburg, Germany}

\author{\\J. L. Ord\'{o}\~{n}ez Carrasco}
\affiliation{Universidad Carlos III de Madrid, avda. Universidad, 30. 28911 Legan\'{e}s (Madrid), Spain}

\author{A. Sinn}
\affiliation{Deutsches Elektronen-Synchrotron DESY, Notkestr. 85, 22607 Hamburg, Germany}

\author{M. Th\'{e}venet}
\email{maxence.thevenet@desy.de}
\affiliation{Deutsches Elektronen-Synchrotron DESY, Notkestr. 85, 22607 Hamburg, Germany}

\date{\today}

\begin{abstract}
The accurate modeling of plasma-based accelerators relies on costly numerical simulations due to the complexity of laser-plasma and beam-plasma interactions.
Several strategies can highly reduce the computational cost compared to 3D first-principles particle-in-cell simulations, such as exploiting the near axial symmetry and quasistatic nature of plasma wakefields in many practical cases.
Here, we propose a quasistatic algorithm that enables the modeling of axially symmetric plasma wakes without the need of a numerical grid.
The gridless approach allows extremely fine features to be resolved without a dramatic increase in computational cost.
This is critical, e.g., for the design of future plasma-based colliders with nanometer emittance beams.
The proposed model has been implemented in the Wake-T code, where it is coupled to a laser envelope solver and a particle beam pusher to enable the efficient simulation of laser- and beam-driven plasma accelerators.
\end{abstract}

\maketitle


\section{Introduction}
Plasma-based accelerator (PBA) technology has the potential of shrinking the size of future particle accelerators thanks to its capability of sustaining up to \SI{\sim100}{\giga\volt\per\metre} gradients.
The promise of extreme gradients and compactness has driven an increasing interest in the research and development of this technology, with breakthroughs in the achievement of GeV energies~\cite{PhysRevLett.122.084801}, high beam quality~\cite{PhysRevLett.126.104801,PhysRevLett.126.174801,PhysRevLett.126.014801}, reproducibility~\cite{PhysRevX.10.031039}, and even free-electron lasing~\cite{Wang2021, Pompili2022, Labat2023}.

Critical to this progress and to its continuity is the use of numerical simulation tools that enable testing new concepts, optimizing the accelerator performance, and improving the understanding of experimental results.
However, 3D full electromagnetic particle-in-cell (PIC) simulations are computationally demanding.
Due to the Courant-Friedrichs-Lewy (CFL) condition, resolving the sub-micron features of the plasma wake over centimeter to meter distances would result in $\mathcal{O}(10^5)$ to $\mathcal{O}(10^7)$ time steps, respectively, which can require in excess of thousands of node-hours on a high-performance computing cluster~\cite{MYERS2021102833}.
This poses a practical limit to such simulations, and therefore to the development of new solutions to the current challenges of PBA technology.
A particularly challenging example is the start-to-end modeling of plasma-based collider concepts~\cite{lindstrom2021self}, which would require the simulation of tens to hundreds of accelerating stages with extreme resolution due to the nanometer emittance of the accelerated beams.

Developing more efficient algorithms is therefore essential to enable large-scale studies and to agilize the progress of PBAs.
Several approaches have been developed that exploit the symmetries and characteristic properties of plasma wakefield generation to achieve a lower computational cost.
These include the quasistatic approximation (QSA)~\cite{10.1063/1.872134, PhysRevE.53.R2068, PhysRevA.41.4463, PhysRevLett.64.2011}, the transformation into a Lorentz-boosted frame~\cite{PhysRevLett.98.130405}, the use of a laser envelope model~\cite{10.1063/1.872134}, and the use of algorithms with cylindrical or quasi-cylindrical geometry~\cite{LIFSCHITZ20091803}.
A particularly efficient example is the approach proposed by P. Baxevanis and G. Stupakov~\cite{PhysRevAccelBeams.21.071301}, which makes use of the QSA and axial symmetry to develop a model for the plasma response that does not require a numerical grid nor a predictor-corrector loop~\cite{AN2013165}.
However, this method was developed to describe the blowout regime in uniform electron plasmas that are driven by particle beams with analytic density profiles, and did not allow for the modeling of more complex and realistic cases.

Here, we propose a generalization of the gridless algorithm that enables the modeling of arbitrary density profiles, multiple plasma species (including the motion of background plasma ions), as well as a non-uniform radial discretization of the plasma.
The model has been implemented in the Wake-T code~\cite{Ferran_Pousa_2019}, where it has been coupled with a laser envelope solver~\cite{Benedetti_2017} and a particle beam pusher to enable the full simulation of laser- and beam-driven plasma accelerators~\cite{ferranpousa:ipac2023-tupa093}.
While the laser and particle beam solvers do need a numerical grid, the size and resolution of each grid is independent of the others and of the gridless background plasma.
This allows for great flexibility, with grids that can dynamically adapt to the size of the particle beams and that are therefore able to resolve arbitrarily small particle distributions without increasing the number of grid points.
This is critical, among others, for the cost-effective simulation of collider concepts.
Benchmarks of the proposed algorithm against the quasi-3D full-PIC code FBPIC~\cite{LEHE201666} and the 3D QSA code HiPACE++~\cite{DIEDERICHS2022108421} demonstrate the high fidelity and drastic cost reduction of the proposed algorithm.

\section{Gridless quasistatic model}\label{sec:theory}

\subsubsection*{Theoretical framework}

The underlying model for the wakefield generation is based on the gridless approach originally developed in Ref.~\cite{PhysRevAccelBeams.21.071301}, which describes the non-linear, axially symmetric response of a uniform electron plasma with a fixed ion background to the space-charge field of analytical electron beam profiles.
Our model generalizes this initial derivation to non-uniform plasmas with an arbitrary number or mobile species (both electrons and ions) which respond to the space-charge fields of beam particle distributions as well as the ponderomotive force of a laser pulse.
The result is a numerical model of broad applicability that enables the efficient simulation of laser- and beam-driven plasma accelerators when the wakefield geometry can be assumed to be axisymmetric.

We introduce the model starting from Maxwell's equations in axial symmetry under the QSA~\cite{PhysRevLett.64.2011, 10.1063/1.872134}, using $\zeta = k_p(z - ct)$ and $\rho = k_p r$ as the normalized longitudinal and radial coordinates, where $k_p^{-1}=\sqrt{m_e\epsilon_0c^2/n_0 e^2}$ is the plasma skin depth, $n_0$ is a reference plasma density, $t$ is time, and $z$ and $r$ are the longitudinal and radial dimensions.
This results in the following set of equations:
\begin{subequations}\label{eqs:maxwell_qsa}
    \begin{align}
        \frac{1}{\rho} \frac{\partial}{\partial \rho} (\rho E_r) + \frac{\partial E_z}{\partial \zeta} &= \sum_{s\in S^{(p)}} q_s n_s + \sum_{s\in S^{(b)}} q_s n_s , \label{eq:maxwell_qsa_1}\\
        \frac{\partial E_z}{\partial \rho} &= \sum_{s\in S^{(p)}} q_s n_s v_{r,s} , \label{eq:maxwell_qsa_2}\\
        \frac{\partial}{\partial \zeta} (B_\theta - E_r) &= - \sum_{s\in S^{(p)}} q_s n_s v_{r,s} , \label{eq:maxwell_qsa_3} \\
        \frac{1}{\rho} \frac{\partial}{\partial \rho} (\rho B_\theta) + \frac{\partial E_z}{\partial \zeta} &=  \sum_{s\in S^{(p)}} q_s n_s v_{z,s} + \sum_{s\in S^{(b)}} q_s n_s ,\label{eq:maxwell_qsa_4}
    \end{align}
\end{subequations}
where the right-hand side contains the contribution of all $N_p$ plasma particle species $S^{(p)}=\{s^{(p)}_i\}_{i\in [1,N_p]}$ and $N_b$ particle beam species $S^{(b)}=\{s^{(b)}_i\}_{i\in [1,N_b]}$ to the charge density and current.
Particle beam species are assumed to be ultra-relativistic and to propagate with no radial motion, i.e., $v_{z,s}=1$ and $v_{r,s}=0$ $\forall s \in S^{(b)}$.
All quantities in Eqs.~(\ref{eq:maxwell_qsa_1}--\ref{eq:maxwell_qsa_4}) are dimensionless:
velocity $\bm{v}$ is normalized to the speed of light $c$, momentum $\bm{u}$ to $m_e c$, mass $m$ to the electron mass $m_e$, charge $q$ to the elementary charge $e$, electric field $\bm{E}$ to $m_e c \omega_p / e$, magnetic field $\bm{B}$ to $m_e \omega_p / e$, scalar potential $\phi$ to $m_e c^2 / e$, vector potential $\bm{a}$ to $m_e c / e$, and particle density $n$ to $n_0$.
For simplicity, we drop the $\zeta$ and $\rho$ dependencies from the notation of any field quantity $\mathcal{F}$, such that $\mathcal{F}(\zeta, \rho)$ is written $\mathcal{F}$ whenever unambiguous.


Using the Lorentz gauge and introducing the potential $\psi = \phi - a_z $ allows the electric and magnetic fields to be written as
\begin{subequations}
    \begin{align}
        E_z &= - \frac{\partial \psi}{\partial \zeta} , \label{eq:lorentz_gauge_1}\\
        B_\theta - E_r &= \frac{\partial \psi}{\partial \rho} , \label{eq:lorentz_gauge_2}
    \end{align}
\end{subequations}
which, if introduced into Eqs.~(\ref{eq:maxwell_qsa_1}--\ref{eq:maxwell_qsa_4}), yields the following equations for $\psi$ and $B_\theta$:
\begin{subequations}
    \begin{align}
        \frac{1}{\rho}\frac{\partial}{\partial \rho} \left( \rho \frac{\partial \psi}{\partial \rho} \right) =   &- \sum_{s\in S^{(p)}} (1 - v_{z,s}) q_s n_s   , \label{eq:phi_b_eq_1}\\
        \frac{\partial}{\partial \rho} \left[ \frac{1}{\rho} \frac{\partial}{\partial \rho}(\rho B_\theta) \right] = &-\frac{\partial}{\partial \zeta} \sum_{s\in S^{(p)}} q_s n_s v_{r,s} \label{eq:phi_b_eq_3}\\
        & + \frac{\partial}{\partial \rho} \left[\sum_{s\in S^{(p)}} q_s n_s v_{z,s} + \sum_{s\in S^{(b)}} q_s n_s\right] . \nonumber
    \end{align}
\end{subequations}
In principle, an additional equation for $\psi$ can be obtained from Eq.~(\ref{eq:maxwell_qsa_2}), i.e, $\frac{\partial}{\partial \rho} \left(\frac{\partial \psi}{\partial \zeta} \right) = \sum_{s\in S^{(p)}} q_s n_s v_{r,s}$, but we will not be making use of it because integrating Eq.~(\ref{eq:phi_b_eq_1}) is already sufficient to find a solution for $\psi$.

The theoretical framework of the wakefield model is completed with the continuity equation for the fluid quantities for each plasma species~\cite{PhysRevAccelBeams.21.071301}
\begin{equation}\label{eq:continuity}
    \frac{1}{\rho} \frac{\partial}{\partial \rho} (\rho n_s v_{r,s}) - \frac{\partial}{\partial \zeta}\left[n_s(1 - v_{z,s})\right] = 0 \ 
\end{equation}
and the constant of motion for an initially cold and unperturbed plasma~\cite{10.1063/1.872134, PhysRevE.53.R2068}
\begin{equation}\label{eq:cold_plasma_constant}
\gamma_s - u_{z,s} + \frac{q_s}{m_s}\psi = 1 \ .
\end{equation}

\subsubsection*{Discretization of plasma species into macroparticles}


Solving Eqs.~(\ref{eq:phi_b_eq_1}) and (\ref{eq:phi_b_eq_3}) requires finding an appropriate expression for the plasma source terms on the right hand side.
The beam density distributions $n_s^{(b)}$ are instead assumed to be known and are provided externally.
They could be, e.g., an analytical expression or a particle density deposited on a grid.
Following the original derivation in Ref.~\cite{PhysRevAccelBeams.21.071301}, the plasma is discretized into a set of $\mathcal{N}_s$ macroparticles $P_s=\{p^s_1, p^s_2, \dots, p^s_{\mathcal{N}_s}\}$ $\forall s \in S^{(p)}$.
Each macroparticle $p^s_j$ is described as a delta function $\delta^s_j(\rho, \zeta) = \delta[\rho-\rho^s_j(\zeta)]$ with a radial and longitudinal position $\rho^s_j(\zeta)$ and $\zeta^s_j$, momenta $u^s_{r,j}(\zeta)$ and $u^s_{z,j}(\zeta)$, and a weight $w^s_j$ that contains the radially integrated particle density represented by the macroparticle.
Similar to the fluid quantities, we will avoid writing the explicit dependency of the macroparticle quantities except were needed.

The macroparticles are initialized at the front of the simulation domain ($\zeta=0$, ahead of all laser and particle beams, where the plasma is unperturbed) with a radial position $\rho^s_{j0} = \rho^s_j (\zeta=0) = \sum_{k<j} \Delta \rho^s_k + \Delta \rho^s_j / 2$, where $\Delta \rho^s_j$ is the radial extent of plasma represented by the $p^s_j$ macroparticle.
In the case of uniform macroparticle spacing (i.e., $\Delta \rho^s_j=\Delta \rho^s$), as considered in Ref.~\cite{PhysRevAccelBeams.21.071301}, the initial radial distribution simplifies to $\rho^s_{j0} = (1/2 + j)\Delta \rho$.
Here, however, we make use of the general expression.
This enables a strategy for refining the plasma discretization transversely only where needed (e.g., close to the axis) and therefore minimizing the computational cost of the algorithm.
This is discussed in detail in Section \ref{sec:adaptive_grids}.
The macroparticle weights depend on the initial radial distribution, and are given by:
\begin{equation}
    w^s_j = \int^{\rho^s_{j0} + \Delta \rho^s_j/2}_{\rho^s_{j0} - \Delta \rho^s_j/2} n_{s}(0, \rho)\rho d\rho \simeq n^s_{j0} \rho^s_{j0}\Delta \rho^s_j ,
    \label{eq:macroparticle_weight}
\end{equation}
where we have assumed that the initial particle density profiles of all plasma species are identical (i.e., $n_s(0,\rho) = n(0,\rho)$ $\forall s \in S^{(p)}$) and that the density is uniform (i.e., $n(0,\rho)=n(0,\rho^s_{j0})=n^s_{j0}$) within the bounds of the integral, i.e., within the immediate surroundings of the macroparticle.

Equations~(\ref{eq:continuity}), (\ref{eq:macroparticle_weight}) and (\ref{eq:cold_plasma_constant}) allow the fluid quantities in the right-hand side of Eqs.~(\ref{eq:phi_b_eq_1}--\ref{eq:phi_b_eq_3}) to be obtained for each species $s\in S^{(p)}$ by summing over all macroparticles $p^s_j \in P_s$:
\begin{subequations}
    \begin{align}
        n_s &= \sum_{p^s_j \in P_s} \frac{w^s_{j}}{\rho^s_j[1 - v^s_{z,j}]}\delta^s_j , \label{eq:fluid_from_mp_1}\\
        n_s (1-v_{z, s}) &= \sum_{p^s_j \in P_s} \frac{w^s_j}{\rho^s_j} \delta^s_j , \label{eq:fluid_from_mp_2}\\
        n_s v_{r, s} &= \sum_{p^s_j \in P_s} \frac{w^s_j v^s_{r,j}}{\rho^s_j[1 - v^s_{z,j}]}\delta^s_j , \label{eq:fluid_from_mp_3}\\
        n_s v_{z, s} &=  \sum_{p^s_j \in P_s} \frac{w^s_j v^s_{z,j}}{\rho^s_j[1 - v^s_{z,j}]} \delta^s_j . \label{eq:fluid_from_mp_4}
    \end{align}
\end{subequations}

\subsubsection*{Equations of motion for plasma macroparticles}

After initialization, the distribution of plasma macroparticles can be perturbed by the space-charge field of particle beams, the ponderomotive force of a laser pulse, as well as the resulting fields generated by the plasma response.
To simplify the notation and avoid the superindex $s$ in the macroparticle quantities, we will use in the following a single index $i$ to identify each plasma macroparticle independently of its corresponding species.
This index is defined such that macroparticles are sorted by increasing radius, i.e., $\rho_{i+1}>\rho_i$ $\forall i$ always.
Assuming that Eq.~(\ref{eq:cold_plasma_constant}) holds, the evolution of each plasma macroparticle is given by the equations of motion
\begin{subequations}\label{eqs:plasma_particle_motion}
    \begin{align}
        \frac{d u_{r,i}}{d\zeta} =& \frac{q_i}{m_i} \left(\frac{\gamma_i}{\bar{\psi}_i}\partial_\rho \psi_i - B_{\theta,i} - \frac{q_i}{4m_i}\frac{\nabla_\perp |\hat{a}|^2_i}{\bar{\psi}_i}\right)  \label{eq:plasma_particle_motion_1}\\
        \frac{d \rho_i}{d\zeta} =& - \frac{u_{r,i}}{\bar{\psi}_i} , \label{eq:plasma_particle_motion_2}\\
        u_{z,i} =& \frac{1 + u_{r,i}^2 - \bar{\psi}_i^2 + \frac{q_i^2}{2m_i^2}|\hat{a}|^2_i }{2\bar{\psi}_i}, \label{eq:plasma_particle_motion_3}\\
        \gamma_i =& \frac{1 + u_{r,i}^2 + \bar{\psi}_i^2 + \frac{q_i^2}{2m_i^2}|\hat{a}|^2_i}{2\bar{\psi}_i}, \label{eq:plasma_particle_motion_4}
    \end{align}
\end{subequations}
where we define $\bar{\psi}_i = 1-\frac{q_i}{m_i} \psi_i$ and  include the ponderomotive force of a linearly polarized laser pulse with a transverse normalized vector potential $a_\perp(\zeta, \rho, t) = \Re[{\hat{a}(\zeta, \rho, t) \, \mathrm{exp}(-ik_0\zeta)]}$.
In Eqs.~(\ref{eq:cold_plasma_constant}--\ref{eq:plasma_particle_motion_4}) and in the following, we adopt the notation $\mathcal{F}_i=\mathcal{F}(\zeta_i, \rho_i)$ and $\partial_\nu \mathcal{F}_i=\frac{\partial \mathcal{F}}{\partial \nu}(\zeta_i, \rho_i) \, \forall \nu \in \{\zeta, \rho\}$ for the field quantities and their derivatives evaluated at the position of the macroparticles.

\subsubsection*{Numerical integration of wakefields}

Equations (\ref{eq:plasma_particle_motion_1}--\ref{eq:plasma_particle_motion_4}) can be integrated numerically by discretizing the longitudinal direction with a finite step $\Delta \zeta$.
Then, by recovering the plasma fluid quantities with Eqs.~(\ref{eq:fluid_from_mp_1}--\ref{eq:fluid_from_mp_4}), a numerical solution of Eqs.~(\ref{eq:phi_b_eq_1}) and (\ref{eq:phi_b_eq_3}) can be obtained.
Starting with Eq.~(\ref{eq:phi_b_eq_3}), the magnetic field can be first separated into two components $B_{\theta} = \bThetaBeamSpecies + \bThetaPlasmaSpecies$ that represent the contribution from the current of the particle beams $\bThetaBeamSpecies$ and the plasma species $\bThetaPlasmaSpecies$.
In this way, the magnetic field from the beams can be obtained from a given charge distribution (either an analytical profile or a discretized density deposited on a grid) as
\begin{equation}
        \frac{\partial}{\partial \rho} \left[ \frac{1}{\rho} \frac{\partial}{\partial \rho}(\rho \bThetaBeamSpecies) \right] = \frac{\partial}{\partial \rho} \sum_{s\in S^{(b)}} q_s n_s \, ,
\end{equation}
while $\bThetaPlasmaSpecies$ is given by
\begin{equation}\label{eq:b_theta_plasma0}
        \frac{\partial}{\partial \rho} \left[ \frac{1}{\rho} \frac{\partial}{\partial \rho}(\rho \bThetaPlasmaSpecies) \right] =
            \sum_{s\in S^{(p)}} q_{s}\left[ \frac{\partial n_{s} v_{z, s}}{\partial \rho}- \frac{\partial  n_{s} v_{r, s} }{\partial \zeta}  \right].
\end{equation}

The radial derivative of the longitudinal plasma current is found using Eqs.~(\ref{eq:cold_plasma_constant}) and (\ref{eq:fluid_from_mp_4})
\begin{equation}\label{eq:dr_jz}
    \sum_{s\in S^{(p)}} q_{s}\frac{\partial n_s v_{z,s}}{\partial \rho} = \sum_{i} \frac{q_i w_i}{\rho_i} \left(\frac{\gamma_i}{\bar{\psi}_i} - 1\right) \partial_\rho \delta_i ,
\end{equation}
while the longitudinal derivative of the radial current can be obtained from Eqs.~(\ref{eq:cold_plasma_constant}), (\ref{eq:fluid_from_mp_3}), (\ref{eq:plasma_particle_motion_1}) and (\ref{eq:plasma_particle_motion_2}) as \cite{PhysRevAccelBeams.21.071301}
\begin{equation}\label{eq:dz_jr}
\begin{split}
    \sum_{s\in S^{(p)}} &q_{s}\frac{\partial n_s v_{r,s}}{\partial \zeta} =
        \sum_{i} \frac{q_i w_i}{\rho_i}
            \Bigg[ \\
                &\frac{q_i}{m_i}\frac{\delta_i}{\bar{\psi}_i}
                    \left(
                         \frac{\gamma_i\partial_\rho \psi_i}{\bar{\psi}_i} - B_{\theta,i}^{(p)} - \frac{q_i}{m_i}\frac{\partial_\rho |a|^2_i}{\bar{\psi}_i}
                    \right) \\
                &+ \frac{q_i}{m_i}\frac{u_{r,i}\delta_i}{\bar{\psi}_i^2}
                    \left(
                        - \frac{u_{r,i}\partial_\rho \psi_i}{\bar{\psi}_i}
                        + \partial_\zeta \psi_i
                    \right)   \\             
                &+ \frac{u_{r,i}^2\partial_\rho \delta_i}{\bar{\psi}_i^2}
                + \frac{u_{r,i}^2\delta_i}{\rho_i\bar{\psi}_i^2}
            \Bigg] .
\end{split}
\end{equation}
This expression is equivalent to the explicit form described in \cite{PhysRevAccelBeams.25.104603}, and therefore allows for a numerical solution of the plasma evolution without the need of a predictor-corrector loop~\cite{HUANG2006658, Mehrling_2014}.

Following our assumption that $\rho_{i+1} > \rho_i$, the magnetic field between macroparticles $i$ and $i+1$ can be expressed as
\begin{equation}\label{eq:b_theta_plasma}
    \bThetaPlasmaSpecies = a_i \rho + \frac{b_i}{\rho} ,
\end{equation}
which follows from the fact that there are no plasma currents in the space between the two plasma macroparticles (delta functions), and where $a_i$ and $b_i$ are obtained by solving the recursive system
\begin{equation}
\label{eq:ai_bi}
\begin{split}
    \begin{pmatrix}
        a_i \\[6pt]
        b_i
    \end{pmatrix}
    =&
    \begin{pmatrix}
        1 + \frac{A_i \rho_i}{2} & \frac{A_i}{2 \rho_i} \\[6pt]
        -\frac{A_i \rho_i^3}{2} & 1 - \frac{A_i \rho_i}{2}
    \end{pmatrix}
    \begin{pmatrix}
        a_{i-1} \\[6pt]
        b_{i-1}
    \end{pmatrix}\\
    &+
    \begin{pmatrix}
        \frac{1}{4}(2B_i + A_i C_i) \\[6pt]
        \frac{1}{4}\rho_i(4C_i - 2B_i \rho_i - A_i C_i \rho_i)
    \end{pmatrix}
\end{split}
\end{equation}
with boundary conditions $b_0=0$ and $a_N=0$~\cite{PhysRevAccelBeams.21.071301}.
The coefficients $A_i$, $B_i$ and $C_i$ result from rewriting Eq.~(\ref{eq:b_theta_plasma}) as $\frac{\partial}{\partial \rho} \left[ \frac{1}{\rho} \frac{\partial}{\partial \rho}(\rho \bThetaPlasmaSpecies) \right] = \sum_i\frac{w_i}{\rho_i}(A_i \delta_i \bThetaPlasmaSpecies + B_i\delta_i + C_i \partial_\rho \delta_i)$ after substituting Eqs.~(\ref{eq:dr_jz}) and (\ref{eq:dz_jr}).
Thus, they are given by
\begin{subequations}
    \begin{align}
        A_i =& \frac{w_i}{\rho_i}\frac{q_i^2}{m_i}\frac{1}{\bar{\psi}_i} , \label{eq:b_coeffs_1} \\
        B_i =&
            - \frac{w_i}{\rho_i} \frac{q_i^2}{m_i} \frac{\gamma_i\partial_\rho \psi_i}{\bar{\psi}_i^2}
            + \frac{w_i}{\rho_i} \frac{q_i^2}{m_i} \frac{B_{\theta,i}^{(b)}}{\bar{\psi}_i} \label{eq:b_coeffs_2} \\
            &+ \frac{w_i}{\rho_i} \frac{q_i^3}{m_i^2} \frac{\partial_\rho |\hat{a}|^2_i}{\bar{\psi}_i^2}
            - \frac{q_i w_i}{\rho_i^2} \frac{u_{r,i}^2}{\bar{\psi}_i^2} \nonumber\\
            &+ \frac{w_i}{\rho_i} \frac{q_i^2}{m_i} \frac{u_{r,i}^2\partial_\rho \psi_i}{\bar{\psi}_i^3}
            - \frac{w_i}{\rho_i} \frac{q_i^2}{m_i} \frac{u_{r,i}\partial_\zeta \psi_i}{\bar{\psi}_i^2} , \nonumber\\
        C_i =& \frac{q_i w_i}{\rho_i} \left(\frac{\gamma_i}{\bar{\psi}_i}-1\right) - \frac{q_i w_i}{\rho_i}\frac{u_{r,i}^2}{\bar{\psi}_i^2} . \label{eq:b_coeffs_3}
    \end{align}
\end{subequations}
Details about how to solve this recursive system of equations can be found in Appendix \ref{app:rec_system}.

A numerical solution for $\psi$ and $\partial\psi/\partial \rho$ can be found by substituting Eq.~(\ref{eq:fluid_from_mp_2}) into Eq.~(\ref{eq:phi_b_eq_1}), which, after integration from the axis to $\rho$, yields
\begin{equation}\label{eq:dpsi_dr}
    \frac{\partial\psi}{\partial \rho} = - \frac{1}{\rho}\sum_{\substack{i \\ \rho_i < \rho}} q_i w_i \, .
\end{equation}

By integrating again Eq.~(\ref{eq:dpsi_dr}) with boundary condition $\psi(\rho\to\infty)=0$ and assuming that the plasma is initially neutral ($\sum_{i} q_{i} w_{i}=0$)  it is found that
\begin{equation}\label{eq:psi}
    \psi = -\sum_{i} q_i w_i \ln (\rho_i) - \sum_{\substack{i \\ \rho_i < \rho}} q_i w_i \ln \left(\frac{\rho}{\rho_i}\right) .
\end{equation}

Finally, taking the longitudinal derivative of Eq.~(\ref{eq:psi}) results in
\begin{equation}\label{eq:dpsi_dzeta}
    \frac{\partial\psi}{\partial \zeta} = - \sum_{i} \frac{q_i w_i}{\rho_i}\frac{u_{r,i}}{\bar{\psi}_i} + \sum_{\substack{i \\ \rho_i < \rho}} \frac{q_i w_i}{\rho_i}\frac{u_{r,i}}{\bar{\psi}_i} ,
\end{equation}
where Eq.~(\ref{eq:plasma_particle_motion_2}) has been used.

\begin{figure}
    \includegraphics[width=\columnwidth]{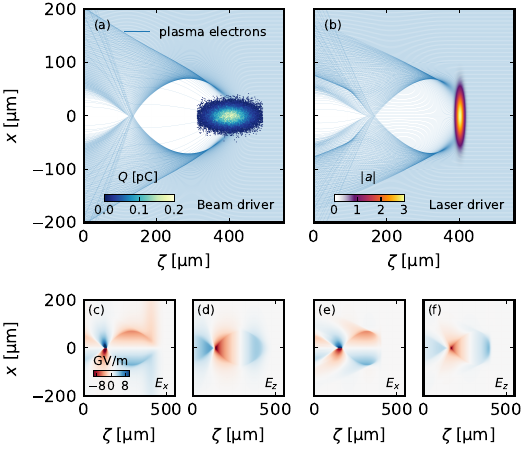}%
	\caption{
        \label{fig:gridless_overview}
        Plasma response to (a) an electron and (b) laser beam as calculated with the gridless quasistatic model.
        The trajectory of each plasma electron macroparticle is shown in blue.
        These trajectories can be used to calculate the resulting electromagnetic fields at any location using Eqs.~(\ref{eq:b_theta_plasma}--\ref{eq:dpsi_dzeta}). 
        In this case, the $E_x$ and $E_z$ fields are shown for both the beam- and laser-driven cases in panels (c)-(f).
    }
\end{figure}

Equations (\ref{eq:b_theta_plasma}--\ref{eq:dpsi_dzeta}) allow the calculation of $\bThetaPlasmaSpecies$, $\psi$, $\partial\psi/\partial\rho$, $\partial\psi/\partial\zeta$ (and therefore of $E_r$ and $E_z$) at any location $\rho$ within the radial plasma column if the position and momentum of the macroparticles, as well the source terms $B^{(b)}_{\theta,i}$, $|\hat{a}|^2_{i}$ and $\partial_\rho|\hat{a}|^2_{i}$ are known.
Thus, they can be used to determine the fields experienced by the plasma macroparticles and to calculate their motion according to Eqs.~(\ref{eq:plasma_particle_motion_1}) and  (\ref{eq:plasma_particle_motion_2}), which can be solved numerically to obtain the response of the plasma to particle and laser beams, as illustrated in Fig. \ref{fig:gridless_overview}.
Numerical algorithms for this task include the Runge-Kutta (used in Ref~\cite{PhysRevAccelBeams.21.071301}) or Adams-Bashforth methods~\cite{butcher2016numerical}, where the latter can be more computationally efficient by not requiring the calculation of intermediate steps.

\subsubsection*{Key algorithmic adjustments}

Special care should be taken when evaluating the fields at the exact location of the plasma macroparticles.
This is due to their shape being a delta function, which results in two main issues that should be handled.

The first one is that the radial sums in Eqs.~(\ref{eq:b_theta_plasma}--\ref{eq:dpsi_dzeta}) present a discontinuity at $\rho=\rho_i$, i.e., where they need to be evaluated to obtain the field values at the location of each plasma macroparticle.
Reference~\cite{PhysRevAccelBeams.21.071301} addresses this issue by taking the average of the two neighboring values, but we have found this approach to still introduce significant error, particularly for the macroparticles closest to the axis.
Instead, we propose to compute the fields as described by the equations above, but using the modified weight $\hat{w}_i = \int^{\rho_{i,0}}_{\rho_{i,0} - \Delta \rho_i/2} n(0,\rho_i) \rho d\rho = w_i/2 - \Delta \rho_i^2/8$ instead of Eq.~(\ref{eq:macroparticle_weight}) for the macroparticle at which the field is being calculated.
The modified weight $\hat{w}_i$, compared to $w_i$, takes into account only the charge of the macroparticle below $\rho_i$, which is the charge that would actually be contributing to the radial integral being computed.

The second issue comes from having to consider multiple plasma species of opposite charge.
This is in contrast to Ref.~\cite{PhysRevAccelBeams.21.071301}, where only the plasma electrons were discretized and a uniform ion background was assumed.
The challenge here is to achieve a field that is radially smooth even though the underlying plasma is a non-smooth distribution of delta functions.
For example, assuming that the same number of electron and ion macroparticles are initialized at exactly the same positions, Eq.~(\ref{eq:dpsi_dr}) would correctly result in $\partial_\rho \psi (\rho)=0$.
However, the tiniest displacement between an electron an its corresponding ion would lead to large, nonphysical gradients between them.
The solution we have adopted to this issue is to solve Eqs.~(\ref{eq:b_theta_plasma}--\ref{eq:dpsi_dzeta}) separately for each plasma species, and afterwards to  gather the fields of a species $s_A$ into another species $s_B$ using linear interpolation.
More specifically, this means that the field $\mathcal{F}_{A}$ (any of $\psi$, $\partial_\rho\psi$, $\partial_\zeta\psi$ or $\bThetaPlasmaSpecies$) of species $s_A$ at the position $\rho_{i,B}$ of a macroparticle from species $s_B$ is given by $\mathcal{F}_{A}(\rho_{i,B}) = \mathcal{F}_{A,j} + \frac{\mathcal{F}_{A,j+1} - \mathcal{F}_{A,j}}{\rho_{j+1,A} - \rho_{j,A}} (\rho_{i,B} - \rho_{j,A})$, where $j$ is the index of the last macroparticle from $A$ with $\rho_{j,A}<\rho_{i,B}$. This method allows for smoother interpolation without resorting to a grid.
Another assumption we make to simplify the algorithm is to compute $\bThetaPlasmaSpecies$ by taking only the electrons into account, since the ion currents are negligible in comparison to those of the electrons due to their much bigger mass ($v_\mathrm{ion} / v_e \propto m_e/m_\mathrm{ion} < 10^{-3}$).

Note that Eqs.~(\ref{eq:b_coeffs_1}--\ref{eq:dpsi_dzeta}) reduce to those in Ref.~\cite{PhysRevAccelBeams.21.071301} in the case of an electron plasma ($q_i=1$, $m_i=1$) with no laser source terms ($|\hat{a}|^2 = 0$) and changing to the longitudinal variable $\xi = -\zeta$.

A remarkable property of this numerical model is that Eqs.~(\ref{eq:b_theta_plasma}--\ref{eq:dpsi_dzeta}) allow the plasma fields to be sampled with arbitrary radial resolution.
Compared to a typical PIC algorithm, the plasma macroparticles do not interact with each other through a numerical grid.
Instead, the fields are calculated directly at the location of each macroparticle, so that the resolution is only limited by the inter-macroparticle distance and the longitudinal step $\Delta\zeta$ of the numerical integrator.
Thus, the presented method can resolve extremely fine features in regions of high macroparticle concentration without the need of a high-resolution grid, and therefore without increasing the computational cost of the algorithm.
This is useful, for example, for modeling the blowout sheath, which is a narrow region where the plasma electrons expelled by the driver accumulate and that features a sharp spike in density and current whose shape is important for determining the blowout fields~\cite{10.1063/1.4775774}.
In this cases, the gridless method is therefore expected to reach convergence at a fraction of the computational cost, compared to an equivalent gridded algorithm.


\section{Integration into Wake-T}

Wake-T (\textbf{Wake}field particle \textbf{T}racker) is an open-source simulation code for charged particle beams that is specialized in the modeling of PBAs~\cite{Ferran_Pousa_2019}.
It aims to provide a fast and inexpensive alternative to general-purpose PIC codes by making use of reduced physical models for the plasma wake.
The approach followed by Wake-T is similar to that of other tools used for the modeling of conventional accelerators, such as ASTRA~\cite{flottmann2011astra}, whose focus is on the numerical integration of the motion of beams of charged particles in the presence of electromagnetic fields.
The fields experienced by the beam particles can be given as analytical expressions or as multidimensional grids whose value is interpolated to the location of the beam particles.
The motion of the beam particles in these fields is integrated using either the Boris method~\cite{boris1970relativistic,Ripperda_2018} or a classical fourth order Runge-Kutta method.
Both of them are parallelized  to take advantage of multi-core CPUs with shared memory (i.e., in a single node).
The code is written in pure \textsc{Python} and uses \textsc{numba}~\cite{10.1145/2833157.2833162} to compile the most compute-intensive functions into fast machine code using \textsc{LLVM}~\cite{1281665}.
Being part of the BLAST toolkit~\cite{blast}, Wake-T is fully compliant with the openPMD data standard~\cite{huebl_2015_33624} for both input and output of the simulations.

\begin{figure*}
    \begin{center}
        \begin{tikzpicture}            \node (start) [etapestart] {Initialize};
            \node (determine) [etape, below=0.7 of start] {Determine next \texttt{element} to evolve \\$(t_e, \Delta t_\mathrm{e})$};
            \node (checktime) [decision, below=1 of determine] {};
            \node (checkbeam) [decision, below=1 of checktime] {};
            \node (checkfield) [decision, below=1 of checkbeam] {};
            \node (checkdiags) [decision, below=1 of checkfield] {};
            \node (terminate) [etapeend, right=1 of checktime] {Terminate};
            \node (evolvebeam) [etape, right=1 of checkbeam] {Evolve \texttt{beam}};
            \node (evolvefield) [etapehighlight, right=1 of checkfield] {Update \texttt{field}};
            \node (evolvediags) [etape, right=1 of checkdiags] {Write \texttt{diagnostics}};
            \node (checktimetext) [checktext, left=0.0 of checktime] {$t_\mathrm{sim}+\Delta t_e>t_\mathrm{max}$};
            \node (checkbeamtext) [checktext, left=0.0 of checkbeam] {Next is \texttt{beam}?};
            \node (checkfieldtext) [checktext, left=0.0 of checkfield] {Next is \texttt{field}?};
            \node (checkdiagstext) [checktext, left=0.0 of checkdiags] {Next is \texttt{diags}?};

            \draw [arrow] (start) -- node[anchor=east] {$t_\mathrm{sim}=0$} (determine);
            \draw [arrow] (determine) -- (checktime);
            \draw [arrow] (checktime) -- node[anchor=east] {no} (checkbeam);
            \draw [arrow] (checkbeam) -- node[anchor=east] {no} (checkfield);
            \draw [arrow] (checkfield) -- node[anchor=east] {no} (checkdiags);
            \draw [arrow] (checktime) -- node[anchor=south] {yes} (terminate);
            \draw [arrow] (checkbeam) -- node[anchor=south] {yes} (evolvebeam);
            \draw [arrow] (checkfield) -- node[anchor=south] {yes} (evolvefield);
            \draw [arrow] (checkdiags) -- node[anchor=south] {yes} (evolvediags);
            \draw [headlessarrow] (evolvebeam) -- ++ (2,0)   node[]{};
            \draw [headlessarrow] (evolvefield) -- ++ (2,0)   node[]{};
            \draw [arrow] (evolvediags) -- ++ (2,0)   node[]{} |- node[align=center,anchor=south east] {$t_\mathrm{sim} = t_e + \Delta t_e$\\$t_e:= t_e + \Delta t_e$} (determine);

            \node (initializepart) [etapestart, right=4.8 of determine] {Initialize plasma macroparticles};
            \node (sort) [etape, below=0.7 of initializepart] {Sort plasma macroparticles radially};
            \node (gather) [etapegrid, below=0.4 of sort] {Gather sources $B^{(b)}_{\theta,i}$, $|\hat{a}|^2_{i}$, $\partial_\rho|\hat{a}|^2_{i}$};
            \node (psi) [etape, below=0.4 of gather] {Calculate $\psi_i$, $\partial_\rho \psi_i$, $\partial_\zeta \psi_i$};
            \node (pz) [etape, below=0.4 of psi] {Calculate $u_{z,i}$ and $\gamma_i$};
            \node (b_theta) [etape, below=0.4 of pz] {Calculate $B_{\theta,i}^{(p)}$};
            \node (b_theta_grid) [etapegrid, right=1 of b_theta] {Calculate $\psi$, $\bThetaPlasmaSpecies$ in external grid};
            \node (checkrho) [decision, above=0.76 of b_theta_grid] {};
            \node (checkendfield) [decision, above=0.76 of checkrho] {};
            
            \node (checkrhotext) [checktext, left=0.0 of checkrho] {Is there a laser?};
            \node (checkendfieldtext) [checktext, left=0.0 of checkendfield] {$\zeta - \Delta\zeta < \zeta_\mathrm{min}$};
            \node (evolvepart) [etape, above=0.76 of checkendfield] {Evolve $\rho_i(\zeta)$, $u_{r,i}(\zeta)$};

            \node (depositrho) [etapegrid, right=1 of checkrho] {Deposit $\chi$ in laser grid};
            \node (endfield) [etapeend, right=1 of checkendfield] {Terminate};

            \node (checklaser) [decision, above=1 of initializepart] {};
            \node (checklasertext) [checktext, left=-0.2 of checklaser] {Is there a laser?};
            \node (evolvelaser) [etape, right=1 of checklaser] {Evolve laser};

            \draw [arrow] (checklaser) ++ (0,0.5)   node[]{} -- (checklaser);
            \draw [arrow] (checklaser) -- node[anchor=south] {yes} (evolvelaser);
            \draw [arrow] (checklaser) -- node[anchor=east] {no} (initializepart);
            \draw [arrow] (initializepart) -- node[anchor=east] {$\zeta=0$}(sort);
            \draw [arrow] (sort) -- (gather);
            \draw [arrow] (gather) -- (psi);
            \draw [arrow] (psi) -- (pz);
            \draw [arrow] (pz) -- (b_theta);
            \draw [arrow] (b_theta) -- (b_theta_grid);
            \draw [arrow] (b_theta_grid) -- (checkrho);
            \draw [arrow] (checkrho) -- node[anchor=east] {no} (checkendfield);
            \draw [arrow] (checkrho) -- node[anchor=south] {yes} (depositrho);
            \draw [arrow] (checkendfield) -- node[anchor=east] {no} (evolvepart);
            \draw [arrow] (checkendfield) -- node[anchor=south] {yes} (endfield);
            \draw [arrow] (evolvepart) |- node[anchor=south east] {$\zeta := \zeta - \Delta\zeta$} (sort);
            \draw [arrow] (evolvelaser.south) |- (initializepart.east);
            \draw [arrow] (depositrho) -- (checkendfield.south);
            \background{initializepart}{checklasertext}{endfield}{b_theta_grid}{bk}
            \draw [dashed, black!50] (evolvefield.south east) -- ([xshift=-0.25em,yshift=-0.25em]bk-a2.north -| bk-a1.east);
            \draw [dashed, black!50] (evolvefield.north east) -- ([xshift=-0.25em,yshift=0.25em]bk-a1.south -| bk-a1.east);

            \node (maillooptitle) [above=0.3 of start, xshift=6em] {Main Wake-T loop};
            \node (maillooptitle) [above=0.3 of evolvelaser, xshift=2em] {Gridless model loop};
            
        \end{tikzpicture}
        
    \end{center}
    \caption{
        \label{fig:gridless_flowchart}
        Simplified view of the main Wake-T particle tracking loop (left) and the implementation of the gridless model (right).
        The main loop evolves each of the \texttt{elements} in the simulation from their initial state up to a maximum final time $t_\mathrm{max}$.
        Since each \texttt{element} can have a different time step,  every iteration begins by determining the next element to evolve by comparing the current simulation time $t_\mathrm{sim}$ with the current time $t_e$ and time step $\Delta t_e$ of each element.       
        When updating the plasma fields with the gridless model, the loop shown on the right is executed.
        The plasma macroparticles are initialized at the right boundary $\zeta=0$ and are iteratively evolved until the left boundary $\zeta_\mathrm{min}$ with a step $\Delta \zeta$.
        Every iteration over longitudinal slices involves sorting the macroparticles radially, computing the required quantities at the location of the macroparticles, and evolving their radial motion.
        A grid is used to communicate between the plasma and the laser and macroparticle beams, as explained in the main text.
        The steps that involve communication with a grid are highlighted with a blue gridded background.
    }
\end{figure*}
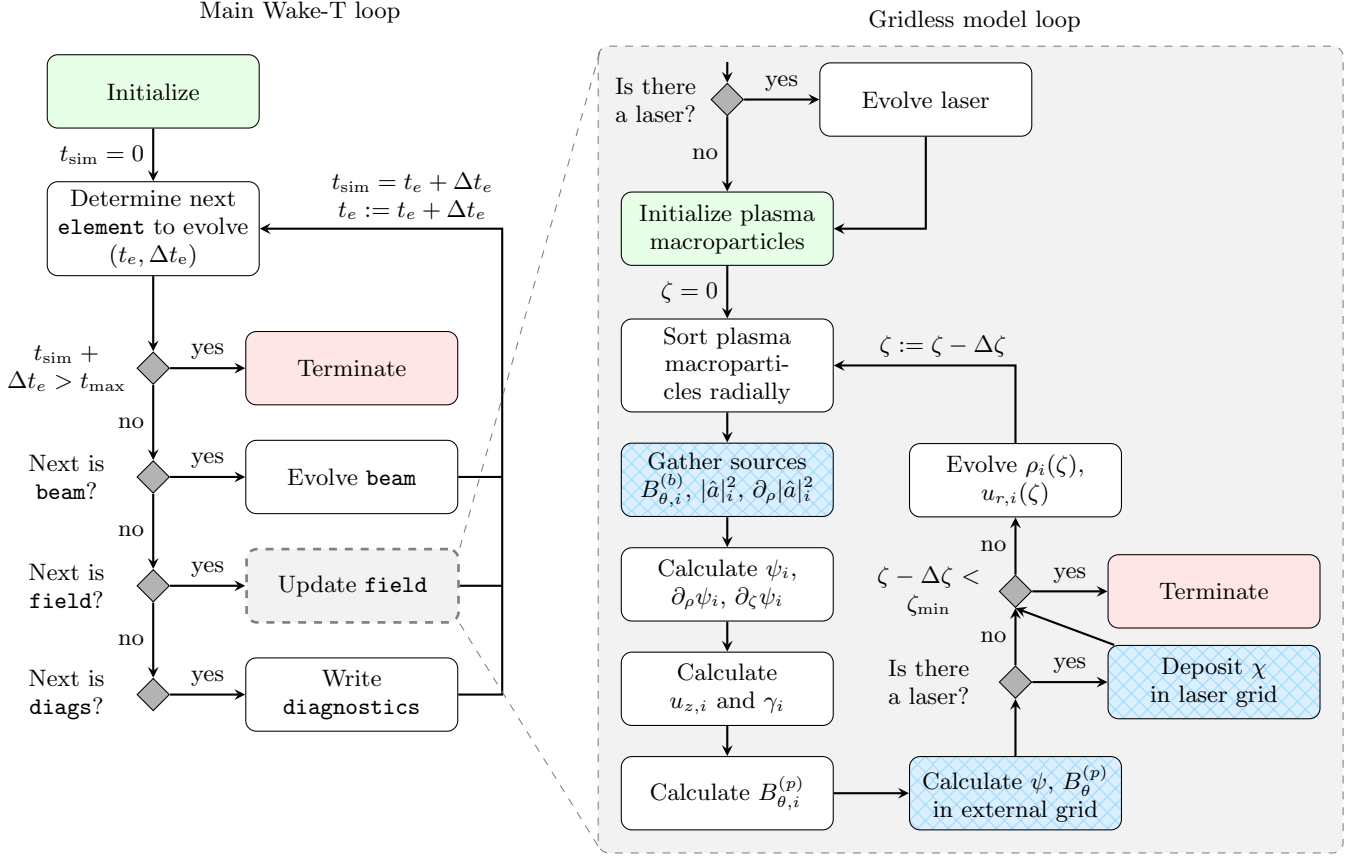

The particle tracking loop in Wake-T evolves the simulation by advancing the beams, the fields, and the output diagnostics in time.
Each of these elements can advance with a different time step, and this time step can be fixed or vary throughout the simulation.
This is done in an iterative process where, at each step, the next \texttt{element} to advance is determined based on the current simulation time, the time step of the elements, and the last time they were advanced. In this context, \texttt{element} is one of \texttt{beam}, \texttt{field} and \texttt{diagnostics}.
This is illustrated in the left diagram of Fig.~\ref{fig:gridless_flowchart} (Main Wake-T loop).

Several reduced wakefield models were previously implemented in Wake-T, including the original gridless~\cite{PhysRevAccelBeams.21.071301} model generalized to support parabolic plasma profiles and a laser envelope solver~\cite{Benedetti_2017}, which allowed for fast optimization of laser-plasma accelerators (LPAs)~\cite{PhysRevAccelBeams.26.084601} or multistage accelerators~\cite{ferranpousa:ipac2023-tupa093}.
Key features enabled by the present work include the use of multiple plasma species (and thus the modeling of ion motion) and arbitrary density profiles.
Furthermore, adaptive grids (see Section \ref{sec:adaptive_grids}) and non-uniform plasma macroparticles spacing enable the modeling of arbitrarily thin beams at constant computational cost.
These features are key, for example, to enable the simulation of future plasma-based collider concepts, such as HAHLF~\cite{Foster_2023}.

A high-level overview of how the new gridless algorithm was implemented in Wake-T is shown in Fig.~\ref{fig:gridless_flowchart}.
At a given time step, the plasma column is iteratively evolved from the front to the back of the domain with a longitudinal step $\Delta\zeta$.
Each iteration starts by sorting the plasma macroparticles radially and gathering the beam and laser source terms at each macroparticle.
This information is then used to calculate the wakefield potential and its derivatives with Eqs.~(\ref{eq:dpsi_dr}--\ref{eq:dpsi_dzeta}).
Knowing $\psi_i$ and the source terms then allows $u_{z,i}$ and $\gamma_i$ to be updated using Eqs.~(\ref{eq:plasma_particle_motion_3}) and (\ref{eq:plasma_particle_motion_4}), and to calculate $B_{\theta,i}^{(p)}$ using Eqs.~(\ref{eq:b_theta_plasma}--\ref{eq:b_coeffs_3}).
Finally, the plasma macroparticles are evolved to the next longitudinal slice (i.e., by $-\Delta\zeta$) using an Adams-Bashforth algorithm of second order.
Once this iterative process is completed ($\zeta = \zeta_{\min}$), the plasma response is known in the whole domain at this time step. Its effect on the laser pulses (refraction) and particle beams (electromagnetic force) can then be accounted for when advancing these by one time step $\Delta t$.


As indicated in Fig.~\ref{fig:gridless_flowchart}, it should be noted that while the algorithm to evolve the plasma macroparticles is gridless, it still needs to be able to communicate with an external grid to handle the laser-plasma and beam-plasma interaction.
The use of a laser grid is required by the envelope solver, which implies that the laser source terms need to be interpolated from the grid onto the plasma macroparticles, and that the macroparticles need to deposit the plasma susceptibility $\chi=n/\gamma$ into the laser grid.
The use of a regular grid for the beam-plasma communication is not strictly required, but it allows for a simpler and potentially more efficient algorithm by using well-established interpolation methods and avoiding the need to sort the beam macroparticles radially and longitudinally at every step.
With this approach, Eqs.~(\ref{eq:b_theta_plasma}--\ref{eq:dpsi_dzeta}) are used to calculate the plasma fields at the grid nodes, and their value at the beam macroparticles is later found through interpolation.
An implementation with direct beam-plasma interaction (i.e., without an intermediate grid) could be explored in the future.
The laser and beam grids are independent from each other and can have different extents and resolutions.
The time step of the laser envelope solver can also be set to be an integer fraction of the time step for updating the fields $\Delta t_\mathrm{laser}=\Delta t_\mathrm{fields}/N_{\mathrm{sub}}$ by performing $N_{\mathrm{sub}}$ subcycles using the same $\chi$.
All of this allows for great flexibility when choosing which parameters to refine in a given simulation.
The longitudinal spacing of the beam grid is always set to match that of the plasma macroparticle solver.

\section{Adaptive grids for particle beams}\label{sec:adaptive_grids}
A potential downside of using an intermediate grid for the beam-plasma communication is that the resolution of the fields seen by the beams is limited by that of the grid.
This would be especially problematic for beams with narrow transverse size $\sigma_r$, as it would require a grid with radial resolution $\Delta r \ll \sigma_r$ and therefore a large number of radial grid points $N_r$ that would increase the cost of the simulation.
This is, for example, one of the main challenges for electromagnetic PIC codes to simulate beams with collider-relevant parameters.
Due to the \si{\nano\metre}-range normalized emittance and \si{\tera\electronvolt}-range energy, $\sigma_x$ would be orders of magnitude smaller than the transverse size of the wake, resulting in an extremely high resolution and, therefore, computational cost.
Fortunately, the gridless method can overcome this issue because the radial extent and resolution of the beam grid is independent of the background plasma.
This allows us to introduce \emph{adaptive grids} (AGs), i.e., grids with a fixed $N_r$ but whose radial and longitudinal extent dynamically adapts to the size of the particle beams.
The number of points per beam width can therefore remain constant, without increasing the simulation cost. This could in principle be implemented per slice.
In order for the fields in the AGs to be accurate, there needs to be sufficient plasma macroparticles in the domain of the AG.
As a general rule, the macroparticle spacing should be smaller than that of the AG, so that the field resolution is not limited by the AG.
Here, we take advantage of a non-uniform initialization of plasma macroparticles enabled by the gridless method presented in Sec.~\ref{sec:theory}: the number of plasma macroparticles can be increased only where needed to keep the computational cost low.
For example, in a simulation with collider-relevant parameters ($\sigma_x\sim\mathcal{O}(\SI{1}{\nano\metre})$) only the region close to the axis would require a higher macroparticle density.

\begin{figure}
    \includegraphics[width=\columnwidth]{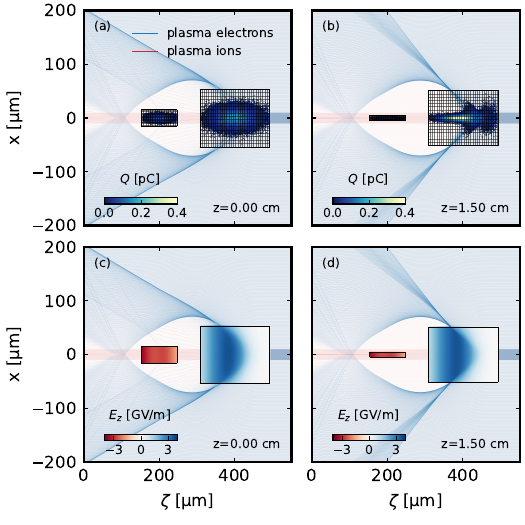}%
	\caption{
        \label{fig:adaptive_grids}
        Use of adaptive grids (AGs) for gathering the plasma fields into the beam macroparticles.
        The grids dynamically adapt to the size of the macroparticle beams at (a) $z=0$ and (b) $z=\SI{1.5}{\centi\metre}$.
        For improved visibility, the grids in (a) and (b) show only a fraction of the grid resolution used in the simulation.
        The AGs allow for high-resolution field sampling only where needed (i.e., in the region the macroparticle beams), as seen in (c) and (d).
        The trajectories of the plasma electrons and ions are shown in blue and red, respectively.
        There, plasma is initialized with more macroparticles close to the axis to capture more accurately the ion motion due to the witness beam.
    }
\end{figure}

An example of using AGs is shown in Fig. \ref{fig:adaptive_grids}, which corresponds to a case with two electron beams (one driver, one witness) in a uniform hydrogen plasma with a density of $\SI{e16}{\per\cubic\centi\metre}$.
Both beams have an initial energy of \SI{1}{\giga\electronvolt} and a relative energy spread of \SI{1}{\percent}.
The driver has a charge of \SI{1}{\nano\coulomb}, an RMS duration of \SI{100}{\femto\second}, a normalized emittance of \SI{10}{\micro\metre} and a transverse size of \SI{10}{\micro\metre}.
The witness beam is centered \SI{200}{\micro\metre} behind the center of the driver and has a charge of \SI{0.5}{\nano\coulomb}, an RMS duration of \SI{50}{\femto\second}, a normalized emittance of \SI{1}{\micro\metre} and a transverse size of \SI{3}{\micro\metre}, which is mismatched to the focusing fields in the plasma.
This causes its transverse size -- and therefore the extent of its AG -- to oscillate throughout the simulation, as observed in Figs. \ref{fig:adaptive_grids}(a) and \ref{fig:adaptive_grids}(b).
The plasma macroparticles have a non-uniform initial distribution, with 8 macroparticles per micron for $r<\SI{10}{\micro\metre}$ and 2 macroparticles per micron in the rest of the domain.
This makes sure that the ion motion arising close to the axis when the witness beam contracts is accurately modeled.

By default, the extent of the AGs is such that all macroparticles in the beam are within the grid.
Optionally, a maximum or a fixed radial extent can be set to prevent them from getting too large if, e.g, a few beam macroparticles escape from the the plasma wake.
Macroparticles that escape the AGs will gather the fields from a lower-resolution base grid covering the full domain of the simulation, just as they would if AGs were not used.

The space charge field of the beams $\bThetaBeamSpecies$ is also contained within their respective AGs, and can be extended beyond them.
If a plasma macroparticle with a radius larger than the extent of the AG needs to gather $\bThetaBeamSpecies$, the beam field at the macroparticle will be given by $B^{(b)}_{\theta, i} = B^{(b)}_{\theta, N_r} r_{N_r} / r_i$, where $r_{N_r}$ and $B^{(b)}_{\theta, N_r}$ are, respectively, the radial position and value of the $\bThetaBeamSpecies$ at the last grid node.

\section{Validation benchmarks}


The accuracy, performance and applicability of the gridless model implemented in Wake-T has been assessed with a laser-driven and a beam-driven benchmarks that are also simulated with the electromagnetic PIC code FBPIC and the 3D quasistatic PIC code HIPACE++, respectively.

The laser-driven setup consists of a plasma stage with a longitudinal density profile given by a \SI{10}{\centi\metre} plateau with an on-axis density  $n_0=\SI{2e17}{\per\cubic\centi\metre}$ and a Gaussian up- and downramp with a length $\sigma_{z,R} = \SI{1}{\milli\metre}$.
The radial profile is parabolic, with a density given by $n(r) = n_0 [ 1 + 4 r^2/(k_p ^2w_0^4)]$.
The laser driver has an energy of \SI{10}{\joule}, a wavelength $\lambda_0=\SI{800}{\nano\metre}$, a peak normalized vector potential $a_0=2.65$, a duration $\tau_\mathrm{FWHM}=\SI{25}{\femto\second}$ and a focal spot size $w_0=\SI{40}{\micro\metre}$ at the entrance to the plateau.
The electron beam is externally injected and has a trapezoidal current profile with a total charge of \SI{114.6}{\pico\coulomb} that has been optimized for beam loading~\cite{katsouleas1987}.
It has a current of \SI{4.26}{\kilo\ampere} (head) and \SI{3.50}{\kilo\ampere} (tail), a head-to-tail duration of \SI{21.2}{\femto\second}, and a time delay between the head and the laser centroid of \SI{184.0}{\femto\second}.
The beam profile is smoothed with a \SI{1}{\micro\metre} Gaussian ramp both at the head and the tail.
It has an initial energy of \SI{200}{\mega\electronvolt} with an RMS relative energy spread of \SI{0.1}{\%}, a normalized emittance of \SI{1}{\micro\metre}  in both planes and a transverse size of \SI{0.92}{\micro\metre} that is matched~\cite{PhysRevSTAB.15.111303} to the focusing fields in the plateau.

The Wake-T simulations were performed using a simulation box with length $L_\mathrm{box}=\SI{120}{\micro\metre}$, where the plasma fields and the laser envelope are updated with a time step $c \, \Delta t_\mathrm{field}=L_\mathrm{box}/2$.
The plasma extends radially up to $2.5w_0$ and the base grid used both for the electron beam and the laser envelope extends up to $4w_0$.
No adaptive grid was needed for the electron beam in this case.
The larger extent of the grid is needed due to the reflective boundary conditions of the envelope solver.
The numerical convergence of the algorithm has been tested by varying the radial and longitudinal resolutions.
In the radial direction, the resolutions $k_p^{-1}/
\Delta r=\SIlist{10;20;40;80}{}$ have been tested while making sure that the underlying gridless plasma features the equivalent of 2 macroparticles per cell.
The longitudinal direction has been discretized using the resolutions $c\tau_\mathrm{FWHM}/\Delta z=\SIlist{10;20;40;80;160}{}$, which also determine the longitudinal step of the plasma particle pusher.
In the FBPIC case, the simulations were performed in the Lorentz boosted frame~\cite{PhysRevLett.98.130405} using $\gamma_\mathrm{boost}=21$, which makes the full plasma stage fit within the boosted simulation window.
The dimensions of the simulation box, the reflective boundary conditions, and the radial resolutions tested were the same as in Wake-T.
In the longitudinal direction, the resolutions used were instead $\lambda_0/\Delta z=\SIlist{10;20;40;80;160;320}{}$ due to the need to capture the laser wavelength.
In all cases, the number of macroparticles per cell was 1, 2 and 4 in the longitudinal, radial and azimuthal direction, respectively.

\begin{figure}
    \includegraphics[width=\columnwidth]{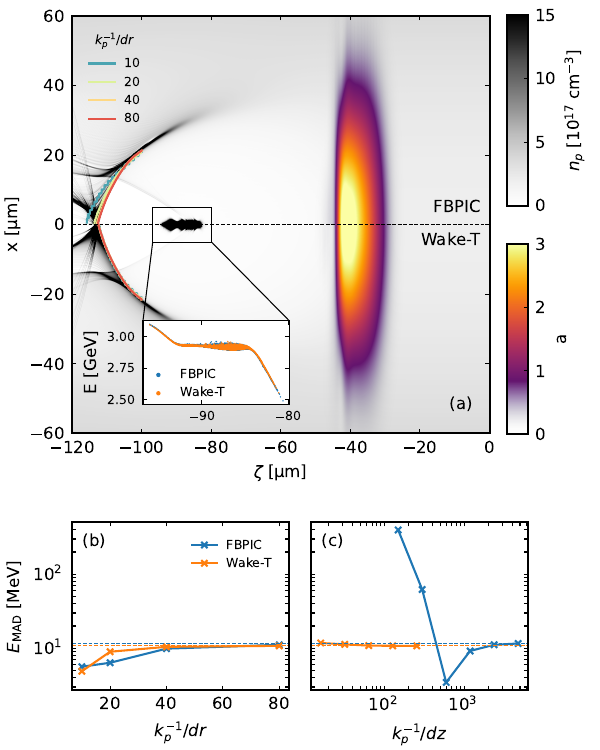}%
	\caption{
        \label{fig:fbpic_comparison}
        Simulation results of the laser-driven study. (a) Side-by-side comparison of the highest-resolution results obtained from FBPIC (top) and Wake-T (bottom) at the same propagation distance in the simulation.
        The colored lines at the back to the wake indicate the shape of the blowout sheath for different radial resolutions.
        The inset shows the final longitudinal phase space of the witness as obtained from both simulation codes.
        The bottom axes show the convergence of the final beam energy spread, measured as the median absolute deviation (MAD), in terms of (b) radial and (c) longitudinal resolution.
    }
\end{figure}

An overview of the simulation results is shown in Fig. \ref{fig:fbpic_comparison}.
The simulations converge to a final witness beam energy and energy spread of \SI{2.91}{\giga\electronvolt} and \SI{0.37}{\%} in Wake-T, and of \SI{2.91}{\giga\electronvolt} and \SI{0.40}{\%} in FBPIC.
The laser evolves considerably in this long plasma stage (the electron beam is accelerated by \SI{2.7}{\giga\electronvolt}), which is well captured by Wake-T.
Wake-T reaches convergence with a longitudinal resolution that is at least a factor 20 lower than in FBPIC.
This can be mainly attributed to the use of laser envelope solver, as expected from Refs.~\cite{Massimo_2019,10.1063/5.0050580}.
The radial convergence is also faster in Wake-T, as indicated by the final beam parameters and by the shape of the blowout sheath highlighted in Fig.~\ref{fig:fbpic_comparison}~(a).
While the sheath at the back of the wake appears identical in all Wake-T simulations, its shape changes in FBPIC unless the resolution is increased above $k_p^{-1}/\Delta r=80$.
The faster radial convergence is likely a feature of the gridless model, which can naturally resolve regions of  high radial macroparticle density.
As a result of the faster convergence and simpler physical model, the converged Wake-T simulations (using $k_p^{-1}/\Delta r=40$ and $c\tau_\mathrm{FWHM}/\Delta z=80 \implies k_p^{-1}/\Delta z \sim 127$) can be performed in only \SI{7}{\minute} on a single CPU core, while an FBPIC simulation of comparable convergence level (using $k_p^{-1}/\Delta r=80$ and $\lambda_0/\Delta z=160 \implies k_p^{-1}/\Delta z \sim 2376$) would require \SI{9.8}{\hour} making full use of an NVIDIA A100 GPU.
This considerable reduction in computational cost and runtime enables, e.g., fast prototyping of LPA design, simulation of long multi-stage setups~\cite{ferranpousa:ipac2023-tupa093}, large-scale optimization, and stability and jitter studies.

The beam-driven benchmark consists of a 4-meter-long lithium stage with a uniform density of \SI{e16}{\per\cubic\centi\metre} preionized to the first level.
The choice of lithium is motivated by the large energy difference between the first and second ionization levels, which prevents a narrow, high-intensity witness beam with collider-like parameters from further ionizing the plasma.
The driver beam is initially Gaussian in all directions and has an energy of \SI{20}{\giga\electronvolt}, an energy spread of \SI{1}{\%}, a charge of \SI{2}{\nano\coulomb}, an RMS duration $\sigma_t=\omega_p^{-1}\simeq \SI{177.3}{\femto\second}$, a normalized emittance of \SI{10}{\micro\metre} and a matched transverse size of \SI{1.95}{\micro\metre}.
The witness has a longitudinal trapezoidal profile that, as in the laser-driven case, has been optimized for beam loading.
It has a total charge of \SI{833}{\pico\coulomb}, a head-to-tail duration of \SI{217.3}{\femto\second}, a current ratio of $0.32$ between the tail and the head, and a time delay between the head and the driver centroid of \SI{694.9}{\femto\second}.
The initial energy is \SI{175}{\giga\electronvolt} with an energy spread of \SI{0.35}{\%}.
The normalized emittance is \SI{135}{\nano\metre} in both planes, with a matched transverse size of \SI{132}{\nano\metre}.
These parameters are in the range of what could be expected from an acceleration stage identified in the European Strategy for Particle Physics roadmap~\cite{adolphsen2022european}, or a plasma-based Higgs factory, such as the HALHF proposal~\cite{Foster_2023} if round beams were used~\cite{diederichs2024resonant}.


Due to the small transverse size of the witness, extreme scale discrepancies make this scenario very challenging even with the most advanced electromagnetic PIC codes like FBPIC, and the use of AGs in Wake-T and of mesh refinement (MR) in HiPACE++~\cite{NIC.MR,hipacemr} is required to improve the efficiency and reduce the run time.
In order to assess the advantages of AGs, Wake-T simulations with and without AGs have been performed.
Similarly, ion motion is expected to play a significant role due to the high intensity of the witness.
Thus, simulations with and without mobile ions are included in the study.
The simulation box in Wake-T has a length $L_\mathrm{box}=\SI{650}{\micro\metre}$ and a plasma column that extends up to $r_\mathrm{max}=\SI{300}{\micro\metre}$.
The longitudinal resolution of all grids and of the plasma pusher is $L_\mathrm{box} / \Delta z = 600$, while the radial resolution is varied to perform a convergence study.
The time step for updating the plasma wakefields is $c \Delta t_\mathrm{field}=\SI{2}{\milli\metre}$.
When not using AGs, both driver and witness share the same base grid, which has the same radial extent as the plasma.
The radial resolutions tested for this case were $r_\mathrm{max}/\Delta r=\SIlist{256;512;1024;2048;4096;8192;16384;32768}{}$.
When making use of AGs, each beam gets an associated grid with a radial extent $r_{\mathrm{max},w}=\SI{1}{\micro\metre}$ for the witness and $r_{\mathrm{max},d}=\SI{20}{\micro\metre}$ for the driver.
The radial resolutions tested for the witness AG were $r_{\mathrm{max},w}/\Delta r_w=\SIlist{4;8;16;32;64;128;256}{}$; and $r_{\mathrm{max},d}/\Delta r_d= 16 r_{\mathrm{max},w}/\Delta r_w$ for the driver.
The radial size of both AGs is fixed throughout the computation to keep $\Delta r_w$ and $\Delta r_d$ constant and therefore simplify the analysis of the convergence study.
The macroparticle distributions of both beams are also identical in all simulations, having \num{8e5} and \num{4e5} macroparticles for the driver and witness, respectively.
Although this number highly exceeds the requirements of the lower resolution simulations (which could therefore be proportionally less expensive), using the exact same distribution allows the convergence scan to exhibit much smoother behavior.
Otherwise, the behavior of the beam parameters would be hindered by differently random beam distributions.

The 3D simulations with HiPACE++ were performed on the JUWELS supercomputer~\cite{JUWELS} using the same box length and longitudinal resolution as in Wake-T, and with a radial extent of \SI{300}{\micro\metre}.
The transverse resolution when not using MR was varied between $r_\mathrm{max}/\Delta r=\SIlist{256;512;1024;2048;4096}{}$.
The cases with MR used a base grid with $r_\mathrm{max}/\Delta r_0=\SIlist{4096}{}$ and an on-axis refined patch extending 7 radial cells of the base grid with a resolution of $r_\mathrm{max}/\Delta r_1=\SIlist{4096;8192;16384;32768;65536;131072}{}$.
Note that HiPACE++ resolutions are discussed in terms of radius for direct comparison with Wake-T: for the same resolution, HiPACE++ uses twice as many points in $x$ and $y$ as Wake-T does in $r$. 
One macroparticle per cell was used in all dimensions, both in the base grid and in the refined patch.
As in Wake-T, the same driver and witness beams were used across all simulations, both with \num{e8} macroparticles.

\begin{figure}
    \includegraphics[width=\columnwidth]{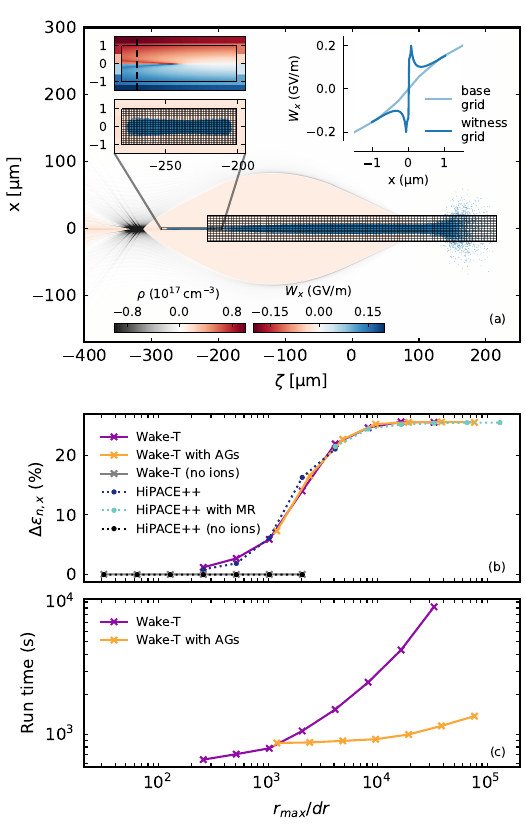}%
	\caption{
        \label{fig:hipace_comparison}
        (a) Simulation of a beam-driven plasma stage with collider-relevant parameters performed with Wake-T.
        The use of adaptive grids (AGs) is advantageous here due to the small transverse size of the witness beam.
        The insets in (a) show a detailed view of the witness beam and its AG.
        The fields in the AG feature a much higher resolution than in the rest of the domain, and are able to resolve the ion motion cone.
        A transverse lineout of $W_x = E_x - B_y$ in the base grid and the AG is shown in the top right inset.
        (b) Final emittance growth for different effective radial resolution as obtained from Wake-T (with and without AGs) and HiPACE++ (with and without MR), including a case where ion motion is disabled.
        (c) Comparison of the total runtime of the Wake-T simulations with and without AGs.
    }
\end{figure}

The results of the simulation study are summarized in Fig. \ref{fig:hipace_comparison}.
The witness beam experiences an energy gain of \SI{17.1}{\giga\electronvolt} and an emittance growth of \SI{25}{\%}, where the latter occurs due to the strong ion motion caused by the witness beam itself.
Both Wake-T and HiPACE++ converge at the same rate and to the same result.
This confirms that the ion motion is correctly modeled in Wake-T, and therefore the validity of the multi-species implementation in the presented gridless model.
Another highlight is that the use of AGs in Wake-T and of MR in HiPACE++ allows for convergence to the same results while allowing for strong reduction of the computational cost.
The use of AGs allows the electron beams to experience the fields with high resolution (see Fig. \ref{fig:hipace_comparison}(a)), reaching convergence with simulations that are an order of magnitude faster (\SI{25}{\minute} vs \SI{2.8}{\hour}) for the same equivalent resolution.
It should be noted that the cost of the Wake-T simulations is artificially dominated by the beam particle pusher, due to the large number of beam macroparticles.
Since convergence is already reached with $r_\mathrm{max}/\Delta r\sim\num{20000}$, which is 4 times smaller than the highest resolution tested, the number of beam macroparticles could also be safely reduced by the same factor.
Thus, when adjusting the number of beam macroparticles accordingly, a converged simulation can be performed in \SI{10}{\minute} on a single CPU core.
The high efficiency demonstrated by the gridless model therefore enables the study of future plasma-based collider concepts with accurate, axisymmetric simulations of the plasma wake.

\section{Conclusion}

The presented gridless quasistatic model makes it possible to simulate axially symmetric plasma wakes with high fidelity at a reduced cost.
Our implementation of the model in the Wake-T code is coupled to a laser envelope solver and a particle beam pusher, and proves to be an accurate and cost-effective alternative to typical PIC codes.
In a laser-driven case, it converges to equivalent results with a resolution that is at least a factor $\sim 40$ lower (factor $\sim 20$ in longitudinal direction and factor $\sim 2$ in radial direction) than with the cutting-edge electromagnetic PIC code (FBPIC), which uses the most advanced numerical schemes, thus reducing the computation runtime from \SI{9.8}{\hour} on an NVIDIA A100 GPU to \SI{7}{\minute} on a single CPU core.
Another advantage of the gridless algorithm is that it allows for a straightforward implementation of local refinement of the field resolution by initializing more plasma macroparticles in the region of interest (e.g., close to the axis), and is devoid of numerical artifacts on the refined grid boundaries.
This keeps the computational cost down while allowing for a high effective resolution.
Although the plasma solver is in itself gridless, the current implementation makes use of an intermediate regular grid for the communication between the plasma and the laser and particle beams.
With this approach, the plasma fields are first precomputed in the regular grid and then interpolated onto the beam macroparticles.
In order to keep the advantages of local field refinement with this approach, the concept of adaptive grids has been introduced.
These grids dynamically adapt to the size of the beams so that the fields are only sampled where needed while keeping a constant resolution relative to the beam size.
Simulations with and without adaptive grids demonstrate convergence to the same solution, and are in full agreement with the results obtained with the 3D quasistatic PIC code HiPACE++.
Furthermore, the use of adaptive grids together with on-axis refinement yields significant savings in computational time (factor $\sim10$ in the presented study) when simulating extremely narrow beams, such as those needed for collider applications (\si{\nano\metre}-level emittance).
As an example, a $\SI{\sim20}{\giga\electronvolt}$, $\SI{4}{\metre}$-long plasma stage with collider-relevant parameters can be simulated in \SI{10}{\minute} on a single CPU core.
This enables simulation studies that would otherwise be prohibitively expensive with current tools, and which are critically needed to solve the challenges of plasma acceleration towards becoming a viable technology for future high-energy colliders.


\section*{Acknowledgment}
We thank P. Baxevanis and G. Stupakov for useful discussions and for providing access to the PLEBS code;
R. Lehe, A. Huebl and J.-L. Vay for fruitful discussions regarding the gridless algorithm;
and C. Benedetti for advice in the implementation of the laser envelope model in Wake-T.
This research was supported in part through the Maxwell computational resources operated at DESY, Hamburg, Germany.
The authors gratefully acknowledge the Gauss Centre for Supercomputing e.V. (www.gauss-centre.eu) for funding this project by providing computing time through the John von Neumann Institute for Computing (NIC) on the GCS Supercomputer JUWELS~\cite{JUWELS} at Jülich Supercomputing Centre (JSC).
This work was funded by the Deutsche Forschungsgemeinschaft (DFG, German Research Foundation)—491245950.
No large language models (LLMs) or other generative-AI tools were used in the writing of this manuscript.

\section*{Author Contributions}
A. Ferran Pousa wrote the manuscript, created the figures, developed and implemented the gridless quasistatic model, and carried out the Wake-T and FBPIC simulations.
W. M. den Hertog implemented the laser envelope model in Wake-T.
S. Diederichs performed the HiPACE++ simulations.
A. Martinez de la Ossa contributed to the writing of the manuscript.
J. L. Ord\'{o}\~{n}ez Carrasco implemented the Boris particle pusher in Wake-T.
A. Sinn contributed to the implementation of the gridless quasistatic model in Wake-T.
M. Th\'{e}venet contributed to the writing of the manuscript and the derivation of the gridless quasistatic model.

\appendix

\section{Solve recursive system}\label{app:rec_system}

The recursive system in Eq.~(\ref{eq:ai_bi}) cannot be solved directly, as we would need to know the values of both $a_0$ and $a_N$ (or, alternatively, of $b_0$ and $b_N$).
Instead, we have the conditions $a_N=0$ and $b_0=0$, which are at opposite ends of the iteration.
To solve this, Eq.~(\ref{eq:ai_bi}) can be rewritten so that $a_i$ and $b_i$ are linear functions of $a_0$
\begin{subequations}
    \begin{align}
        a_i &= K_i a_0 + T_i , \label{eq:app1_ai}\\
        b_i &= U_i a_0 + P_i , \label{eq:app1_bi}
    \end{align}
\end{subequations}
where the coefficients $K_i$, $U_i$, $T_i$, $P_i$ are determined by solving the recursive systems
\begin{equation}
    \begin{pmatrix}
        K_i \\[6pt]
        U_i
    \end{pmatrix}
    =
    \begin{pmatrix}
        1 + \frac{A_i r_i}{2} & \frac{A_i}{2 r_i} \\[6pt]
        -\frac{A_i r_i^3}{2} & 1 - \frac{A_i r_i}{2}
    \end{pmatrix}
    \begin{pmatrix}
        K_{i-1} \\[6pt]
        U_{i-1}
    \end{pmatrix}
\end{equation}
and
\begin{equation}
\begin{split}
    \begin{pmatrix}
        T_i \\[6pt]
        P_i
    \end{pmatrix}
    =&
    \begin{pmatrix}
        1 + \frac{A_i r_i}{2} & \frac{A_i}{2 r_i} \\[6pt]
        -\frac{A_i r_i^3}{2} & 1 - \frac{A_i r_i}{2}
    \end{pmatrix}
    \begin{pmatrix}
        T_{i-1} \\[6pt]
        P_{i-1}
    \end{pmatrix}\\
    &+
    \begin{pmatrix}
        \frac{1}{4}(2B_i + A_i C_i) \\[6pt]
        \frac{1}{4}r_i(4C_i - 2B_i r_i - A_i C_i r_i)
    \end{pmatrix}
\end{split}
\end{equation}
which, in this case, do have a set of initial conditions $K_0=1$, $U_0=0$, $T_0=0$, $P_0=0$ all at $i=0$.

Once the system of equations has been solved iteratively to obtain the value of $K_N$ and $T_N$, we have that $a_N = K_N a_0 + T_N$.
Since $a_N=0$, this implies that
\begin{equation}
    a_0 = -\frac{T_N}{K_N} , \label{eq:app1_a0}
\end{equation}
which allows $a_i$ and $b_i$ to be obtained from Eqs.~(\ref{eq:app1_ai}) and (\ref{eq:app1_bi}), and therefore to calculate $\bThetaPlasmaSpecies$ according to Eq.~(\ref{eq:b_theta_plasma}).

\bibliography{references}

\end{document}